\pdfoutput=1
\documentclass[12pt]{article}

\usepackage[utf8]{inputenc}

\usepackage{amsmath,latexsym,amssymb}

\usepackage[colorlinks=true,citecolor=magenta,urlcolor=blue]{hyperref}

\usepackage{graphicx}

\usepackage{geometry}
\newgeometry{vmargin={20mm}, hmargin={20mm,20mm}}

\newcommand{\R}{\mathbb{R}}

\usepackage{subcaption}

\begin{document}

\begin{center}

{\Large{\underline{\bf En memoria de Peter Higgs (1929 -- 2024)}}} \vspace*{2mm} \vspace*{8mm} \\

{\small José Antonio García-Hernández y Wolfgang Bietenholz \\
  Instituto de Ciencias Nucleares \\
  Universidad Nacional Autónoma de México (UNAM) \\
Apartado Postal 70-543, 04510 Ciudad de México, México}
 \vspace*{6mm} \\

\end{center}

\noindent
Peter Higgs fue un físico teórico británico, famoso por su
trabajo de 1964 donde propuso un mecanismo que puede generar masas
para partículas elementales, conforme a la simetría de norma.
Medio siglo más tarde, dos experimentos del CERN confirmaron
que este mecanismo está realizado en la naturaleza.
El 8 de abril nos llegó la triste noticia del fallecimiento del
gran pionero de la fisica de partículas elementales.
Este artículo es dedicado a su memoria, así como al mecanismo
y a la partícula que llevan su nombre.

\section{Datos biogr\'aficos y contexto hist\'orico}

Peter Higgs nació en 1929 en Newcastle, Inglaterra, por lo que
pasó su juventud parcialmente durante la Segunda Guerra Mundial,
circunstancia que complicó un poco su formación escolar.
Después de la guerra, estudió en Londres, primero matemáticas,
luego física. En 1954, con sólo 25 años terminó su
doctorado en el {\it King's College.}

Luego trabajó temporalmente en la Universidad de Edimburgo,
en el {\it University College} y en el {\it Imperial College,} ambos
en Londres.
En 1960 regresó a Edimburgo -- ciudad que le encantó y donde
había llegado por primera vez en 1949, como estudiante viajando
con {\it auto-stop} -- para ocupar un puesto de catedrático y
quedarse allí toda su vida.

En 1964, a los 35 años, escribió sus dos artículos famosos
(y otro sobre el mismo tema en 1966) \cite{Higgs} que llamaron la
atención y condujeron a invitaciones para presentar seminarios en
Princeton y Harvard en 1966. Tenía que tratar con audiencias críticas,
Sidney Coleman comentó más tarde que en Harvard ``querían
romper en pedazos al idiota que pensaba que podía evadir el
Teorema de Goldstone'' \cite{boson}. Resultó que su concepto siguió
de pie, pero aún sin aplicación fenomenológica
(sus artículos trataron con un modelo de juguete). Además,
se enteró por Yoichiro Nambu (el árbitro de uno de sus
artículos) de un trabajo parecido \cite{EB}, publicado
15 días antes del primer artículo de Higgs sobre el tema.
Los autores eran Fran\c{c}ois Englert y Robert Brout, quienes
trabajaban en Bruselas, Bélgica. Dos meses después apareció
otro artículo relacionado, escrito en Londres por
Gerald Guralnik, Carl Hagen y Tom Kibble \cite{GHK}, pero ellos
conocieron y citaron los trabajos anteriores de Englert, Brout y Higgs.

El mecanismo que estos tres artículos propusieron no era
totalmente nuevo: había sido establecido en 1962/3 en el contexto de la
materia condensada, por Philip Anderson \cite{Anderson}. Él aplicó
conceptos de Julian Schwinger \cite{Schwinger} para la explicación
teórica de la masa de una partícula de norma a la teoría de
superconductores. Englert, Brout y Higgs presentaron una
extensión a modelos relativistas.

\begin{figure}[h]
	\centering
	\begin{subfigure}{0.3\textwidth}
	\includegraphics[scale=0.8]{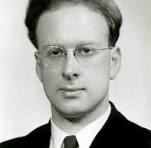}
	\caption{}
	\end{subfigure}
	\begin{subfigure}{0.6\textwidth}
	\centering
	\includegraphics[scale=0.25]{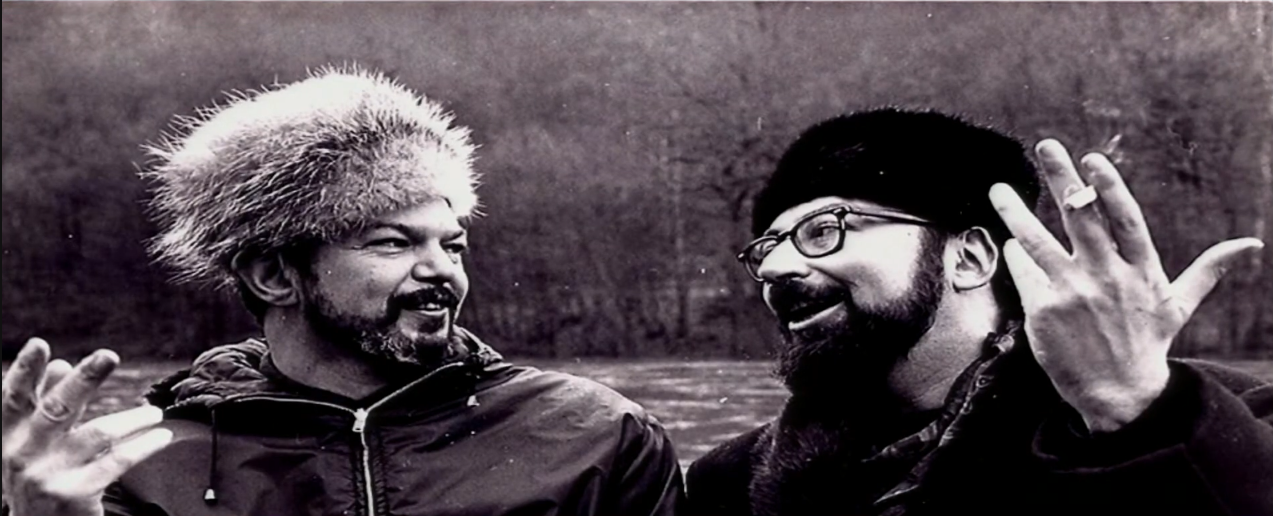}
	\caption{}
	\end{subfigure}
	\caption{(a) Peter Higgs, quien publicó dos artículos breves sobre
        el ahora llamado mecanismo de Higgs en 1964, y otro más extenso
        en 1966. (b) Robert Brout
        (izquierda) y Fran\c{c}ois Englert (derecha). Brout invitó a
        Englert a colaborar en la Universidad de Cornell en 1959 por
        dos años como investigador asociado. Después
        Brout y Englert dejaron Cornell para trabajar en la
        Universidad de Bruselas, Bélgica.}
	\end{figure}
\begin{figure}
\centering
\begin{subfigure}{0.31\textwidth}
	\includegraphics[scale=2]{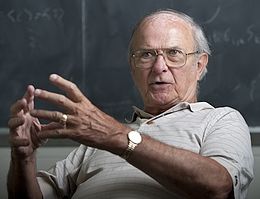}
\end{subfigure}	
\begin{subfigure}{0.31\textwidth}
	\includegraphics[scale=0.5]{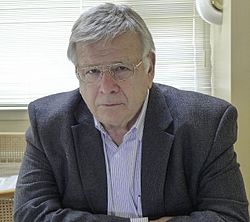}
\end{subfigure}	
\begin{subfigure}{0.31\textwidth}
	\includegraphics[scale=0.25]{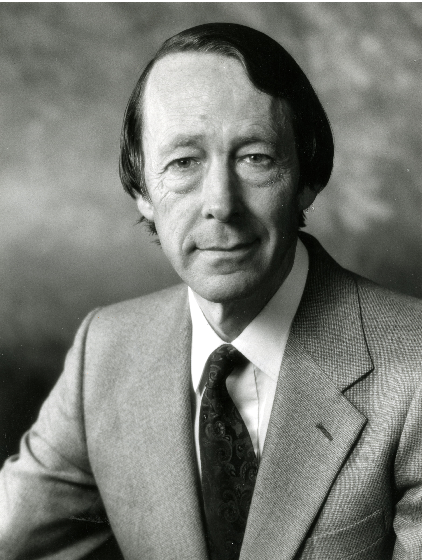}
\end{subfigure}	

	\caption{Los otros descubridores del mecanismo de Higgs.
        De izquierda a derecha, Carl Hagen, Gerald Guralnik y Tom Kibble.} 
\end{figure}

\begin{figure}[h]
\centering
\includegraphics[scale=1]{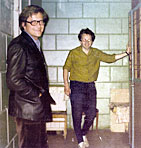}	
\caption{Cuando aún eran {\it teenagers}, Alexander
Migdal (izquierda) y Alexander Polyakov (derecha) descubrieron
el mecanismo de Higgs independientemente del occidente.}
\end{figure}
	\begin{figure}
			\begin{subfigure}{0.5\textwidth}
	\centering	
		\includegraphics[scale=0.15]{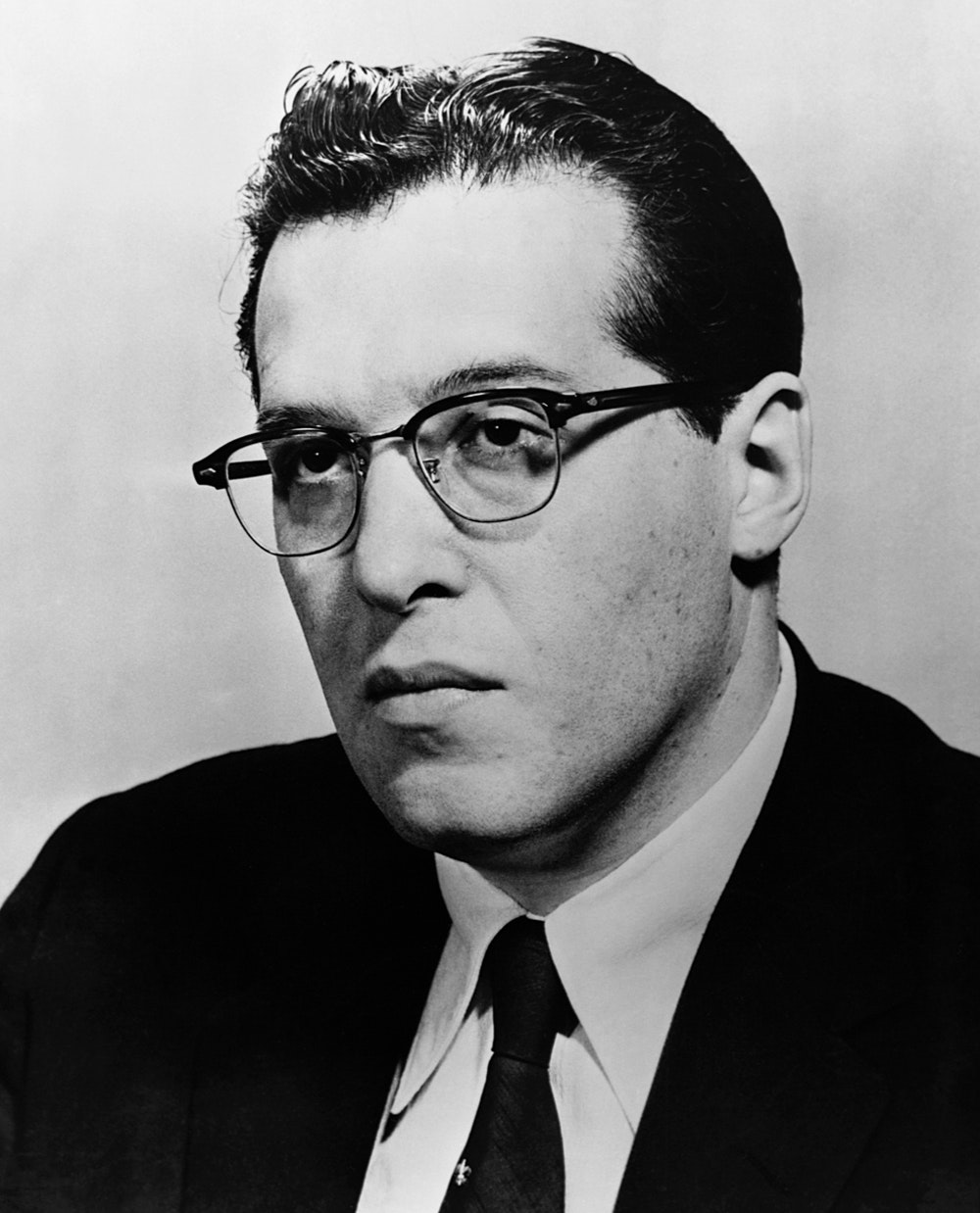}
	\end{subfigure}
	\begin{subfigure}{0.5\textwidth}
	\centering	
\includegraphics[trim = {1cm 1cm 4cm 12cm},clip,scale=0.25]{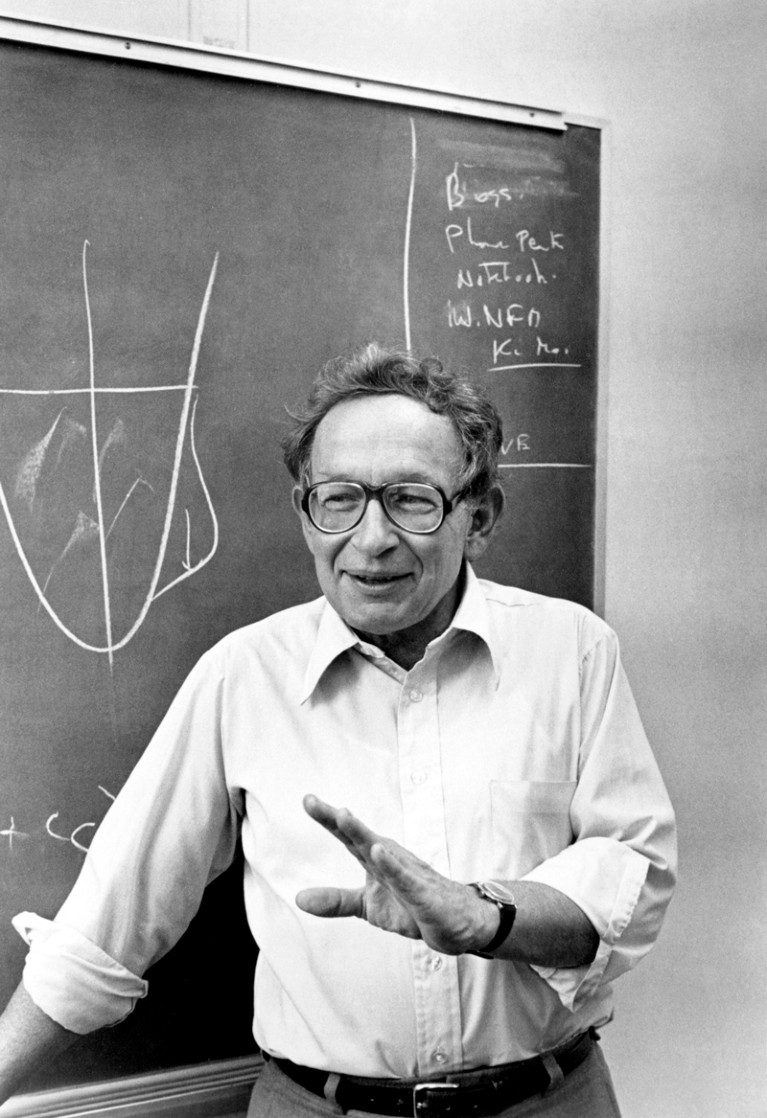}
		\end{subfigure}
\caption{Izquierda: Julian Schwinger, podemos trazar el descubrimiento
del mecanismo de Higgs a su trabajo pionero sobre cómo una simetría
de norma no siempre implicaba un bosón de norma no masivo.
Derecha: Philip Anderson, quien descubrió el mecanismo que da
masa a los bosones de norma en el contexto de la superconductividad.}
	\end{figure}

\begin{figure}[h]
\centering
\includegraphics[scale=0.3]{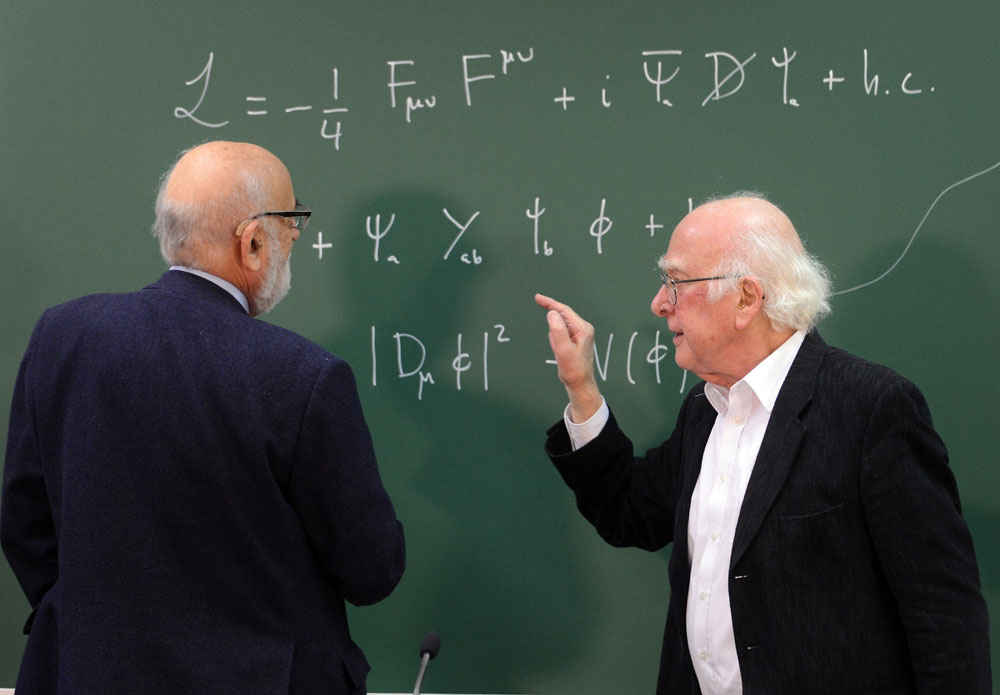}
\caption{Peter Higgs y Fran\c{c}ois Englert discutiendo el lagrangiano
que involucra al ahora llamado campo de Higgs, $\phi$.
?`Qué creen que Higgs le dice a Englert? Tal vez le indique que
en la segunda línea falta una barra en el campo fermiónico
$\bar \psi_a$.}
\end{figure}

Totalmente independientemente, en Moscú en 1964, dos chicos de 19
años, ambos de nombre Alexander (o Sasha), con apellidos Migdal y
Polyakov, discutieron en gran detalle qué significa el rompimiento
de una simetría \cite{Nobel13}. Ellos escribieron otro artículo,
muy diferente pero con conclusiones equivalentes, que fue publicado
en 1966 \cite{MigPol}. En 2010, Migdal visitó México y
relató sobre las dificultades que tuvieron para publicar
este artículo, ya que la comunidad de físicos establecidos
en la Unión Soviética -- bajo el liderazgo de Lev Landau -- rechazó
la teoría cuántica de campos, que todavía era muy controversial
al igual que en el mundo occidental.\footnote{En Alemania
Werner Heisenberg era un oponente influyente contra la teoría cuántica
de campos; su preferencia era el formalismo de la matriz S.}
Finalmente, este artículo fue publicado algo tarde, pero después
ambos Sashas se hicieron famosos por otros trabajos -- en especial,
Polyakov es conocido por descubrir excitaciones topológicas que
llamamos ahora {\em instantones.}

La explicación de cómo  las partículas de norma -- y ciertas partículas
acopladas -- pueden tener masa fue pronto conocida como el {\em mecanismo
de Higgs},\footnote{Consultar la literatura original no conduce a
  una explicación clara sobre el motivo por el que la
  terminología excluye a Brout y Englert.}
  el tema de la Sección 2 de este artículo.
Su aplicación a la fenomenología de partículas elementales 
emergió en 1967/8 por parte de Steven Weinberg \cite{Weinberg} y
Abdus Salam \cite{Salam}.
Ellos integraron este mecanismo al modelo de la interacción
electrodébil que Sheldon Glashow había propuesto en 1961
\cite{Glashow} durante su estancia en Copenhague.
De hecho, Glashow estuvo presente en el seminario de Higgs en Harvard,
y reconoció que era {\it ``a nice model''} \cite{boson}, pero no se le
occurió la idea de que este mecanismo podría ser el remedio
para salvar a su modelo, que él ya había abandonado.

Sin embargo, esta teoría -- ahora conocida como el sector
electrodébil del Modelo Estándar -- aún no era generalmente
aceptada ya que era considerada ``no renormalizable''.
En las teorías cuánticas de campos casi siempre aparecen divergencias
a altas energías, así que requieren una ``regularización'',
una manipulación matemática que convierte las divergencias
en valores finitos. Se dice que una teoría es {\em renormalizable}
si, al final del cálculo, se puede remover la regularización
totalmente y llegar a predicciones finitas para los observables
(esta definición es ligeramente simplista).

La comunidad física cambió su punto de vista en 1971/2, gracias
al trabajo de Gerard 't Hooft, un brillante estudiante de doctorado
en Utrecht, Holanda, quien presentó evidencia a favor de la
renormalizabilidad de dicho modelo (parcialmente junto a
su asesor, Martin Veltman). Estos trabajos \cite{tHooft} causaron
sensación y provocaron un cambio de paradigma en aquella época.

\begin{figure}
\begin{subfigure}{0.5\textwidth}
	\includegraphics[scale=0.5]{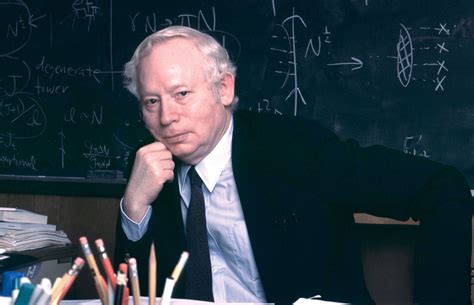}
\end{subfigure}	
\begin{subfigure}{0.5\textwidth}
	\includegraphics[scale=0.5]{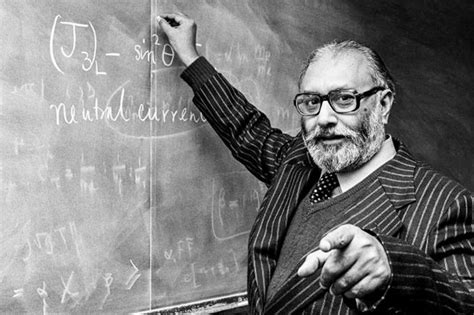}
\end{subfigure}	
\caption{Steven Weinberg (izquierda) y Abdus Salam (derecha)
independientemente integraron el mecanismo de Higgs al sector
electrodébil del Modelo Estándar.}
\end{figure}

\begin{figure}
\centering
\begin{subfigure}{0.5\textwidth}
\centering
\includegraphics[scale=0.45]{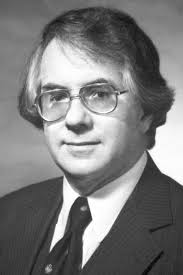}
\end{subfigure}\begin{subfigure}{0.5\textwidth}
\centering
\includegraphics[scale=0.25]{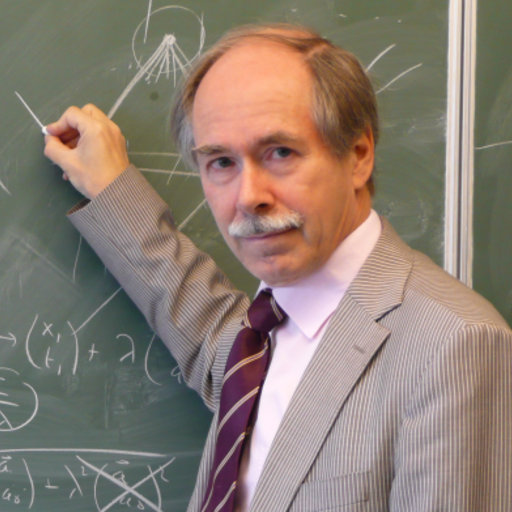}
\end{subfigure}
\caption{Izquierda: Sheldon Glashow, creador de la versión original
de la teoría electrodébil. Derecha: Gerard 't Hooft, famoso por
su trabajo sobre la renormalización del Modelo Estándar,
entre otras logros.}
\end{figure}

La clave para este hito fue un nuevo método, la {\em regularización
dimensional} que está entre los logros principales de la física en
América Latina: fue propuesta primero por dos argentinos, Carlos Bollini
y Juan José Giambiagi en 1971, aunque la publicación \cite{BolGiam}
se demoró hasta 1972. Ellos trabajaron en La Plata, en
circunstancias difíciles durante la dictadura militar \cite{DimReg}.

Agregamos que hoy en día se da menos importancia a la pregunta de
si el Modelo Estándar es renormalizable o no: la tendencia es que se
considera como teoría efectiva y su validez en un gran rango
energético -- que no tiene que extenderse hacia infinito --
es suficiente.

Poco después, en 1973, el Modelo Estándar de las partículas
elementales fue completamente establecido, con un sector
electrodébil \cite{Weinberg,Salam} y otro de la interacción
fuerte \cite{QCD}. El mecanismo de Higgs es indispensable para
proporcionar masas a gran parte de las partículas elementales.
Esto fue una revolución en la física de altas energías como no la
hemos visto más en el medio siglo que siguió, ya que después el
progreso fue relativamente lento.

En el siglo XXI es popular especular sobre física más allá
del Modelo Estándar. Sin embargo, por ahora ninguna de estas propuestas
tiene apoyo sólido de datos experimentales. Por otro lado, los
experimentos han confirmado las predicciones del Modelo Estándar una
y otra vez: muchas veces salen al aire en las noticias que el
Modelo Estándar ha sido ``refutado''
por nuevos resultados, pero al final del análisis, y la repetición
de los experimentos, siempre sus predicciones han
triunfado.\footnote{Como ejemplo reciente, en la segunda parte de la
  década pasada, se difundieron noticias de una tensión entre el Modelo
  Estándar y experimentos con el decaimiento de mesones pesados
  conocidos como ``mesones B''. Al final, esta discrepacia no se
  sustentó. La última moda es el momento magnético del muón,
  para el cual el valor experimental parece un poquito diferente
  del cálculo basado en el Modelo Estándar. Si esto es verdad --
  cosa que no es nada segura -- la predicción se equivoca en un
  nivel relativo de $10^{-10}$: si lo comparamos con la distancia
  entre México y Europa central (Suiza por ejemplo), unos
  10,000\,km, esto corresponde a un posible error de la magntitud
  de milímetros. No obstante, cálculos con simulaciones numéricas
  en la retícula conducen a resultados más cercanos al valor
  experimental \cite{Wittig}, tendencia que indica que incluso
  esta discrepancia
  mínima podría desaparecer con un análisis más preciso,
  igual que todas las supuestas discrepancias anteriores.}
  
El Modelo Estándar es algo incompleto para describir al
universo (faltan por ejemplo la gravitación, la materia oscura y
la energía oscura), pero aún así: se trata de nada menos que la
teoría más precisa y -- en este sentido -- más exitosa en
la historia de la ciencia.\\
 
Higgs ya no participó en este desarollo rápido. Él ya era
tan famoso que podía permitirse casi no publicar más resultados
de investigación a partir de la edad de 40 años.
(En México esto sería un problema serio con el SNII etc.)
 
Fue conocido como una persona tranquila y modesta, casi tímida,
que no buscó la atención mediática o ponerse en el centro de
atención en eventos. Con su mentalidad de abstenerse del
{\it show,} se puede caracterizar como lo opuesto a Feynman.
Esta caracterización corresponde a la impresión de uno de los
autores (WB) quien participó en un congreso en Edimburgo en 1997.
Higgs -- quien era emérito desde 1996 -- apareció en el banquete,
pero muy discreto, simplemente para sentarse en una mesa sin
ningún espectáculo.

Esto no significa que Higgs no tenía convicciones: fue
temporalmente activista por el desarmamento nuclear y
por el movimiento ambiental como miembro de {\it Greenpeace.}\\

Una vez que el Modelo Estándar había sido establecido, su exploración
progresó con trabajo intenso en múltiples países.
En el año 2000, todas sus partículas ya eran encontradas
experimentalmente, menos una: la famosa ``partícula de Higgs'',
involucrada en este mecanismo, como vamos a describir en la Sección 2.

Otra vez, la nomenclatura es tal vez un poco injusta con Englert y
Brout, pero así es la convención de la comunidad.
Higgs no inventó este término (esto lo hizo primero Ben Lee
\cite{boson}), pero también le era incómodo el apodo
absurdo ``partícula de dios'' que no tiene ni el menor sentido.
Este término fue propuesto por la editorial de un libro de divulgación
\cite{Lederman},
obviamente con un objetivo comercial, pero plenamente irresponsable.
Esto motivó que este término se hiciera popular y condujera
a confusión sin fin. Por ejemplo, la iglesia católica de España
llegó a creer que el trabajo del CERN podría tener
algo que ver con teología \cite{Guardian}. !`Debemos
tener cuidado con los términos que usamos!

En el siglo XXI fuimos testigos de una carrera emocionante en la
búsqueda de la partícula de Higgs. En su fase final, era una
competencia entre el Fermilab (cerca de Chicago) y el CERN
(cerca de Ginebra, en una región colindate entre Suiza y Francia).
Después de primeras indicaciones en 2011, en 2012 las
colaboraciones ATLAS y CMS, ambas trabajando de manera independiente
en el Gran Colisionador de Hadrones (LHC) en el CERN, presentaron
evidencia indirecta pero clara para la observación de la
partícula de Higgs, que era tan buscada \cite{ATLASCMS}.

\begin{figure}

\begin{subfigure}{0.5\textwidth}
	\includegraphics[scale=0.27]{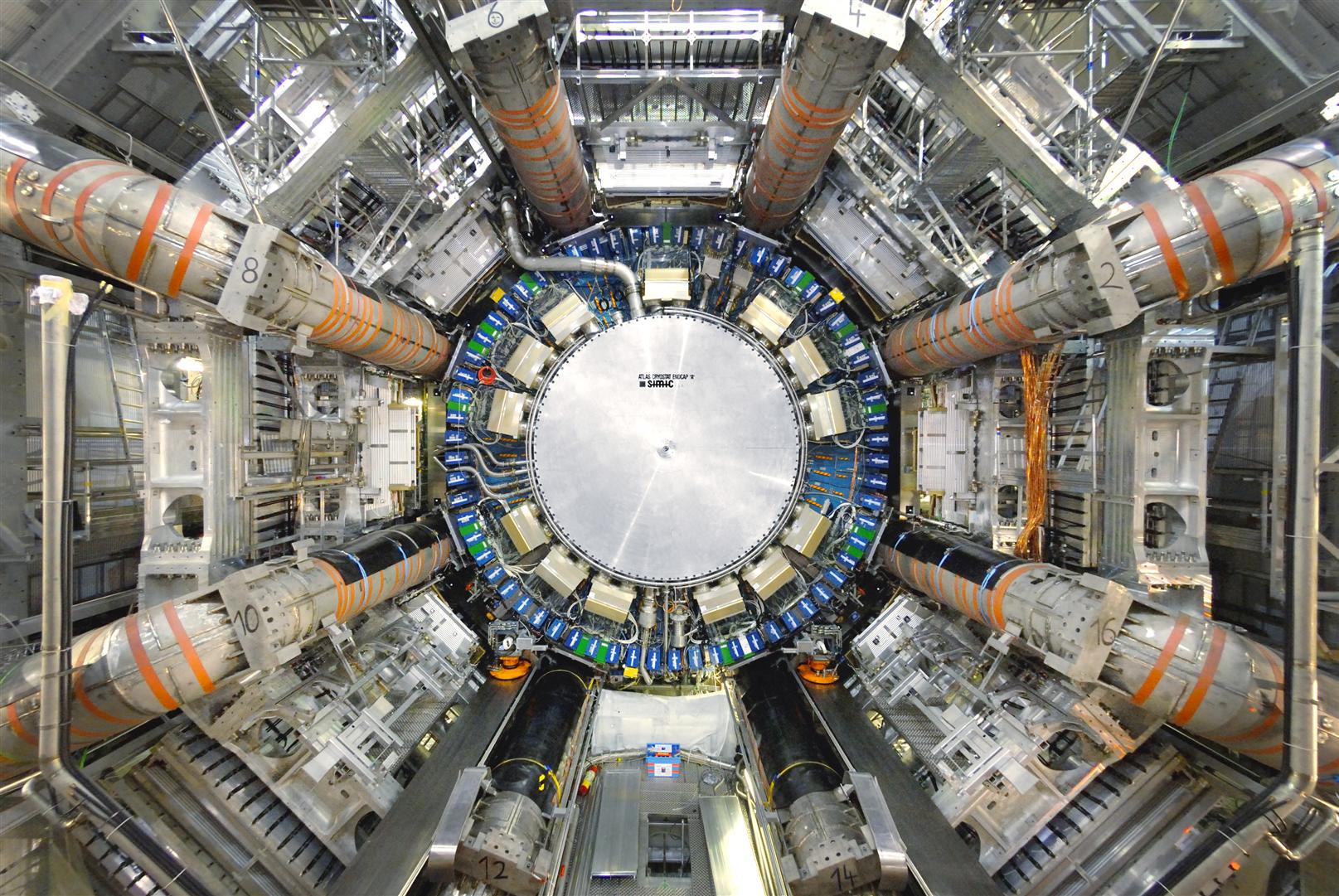}
\end{subfigure}	
\begin{subfigure}{0.5\textwidth}
	\includegraphics[scale=0.9]{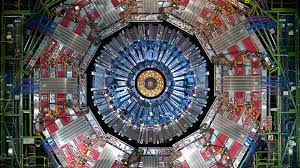}
\end{subfigure}	
\caption{Detectores de la Colaboración ATLAS (izquierda) y CMS
(derecha) en el CERN que confirmaron la existencia del bosón
de Higgs del Modelo Estándar.
Ambos son detectores multipropósito, usados para
analizar colisiones entre partículas de muy alta energía.}
\end{figure}

Con esto, todo el conjunto de partículas del Modelo Estándar
fue observado. Así se confirmó, 48 años después de su
propuesta teórica, que el mecanismo de Higgs está realizado
en la naturaleza. Esto se demoró casi el doble del tiempo
que la observación del neutrino (predicho por Wolfgang Pauli en 1930, y
detectado por Clyde Cowan, Frederick Reines y colaboradores en
1956),\footnote{Ref.\ \cite{neutrinos} revisa la historia y las
propiedades de los neutrinos, desde una perspectiva semi-divulgativa.}
vemos que a veces vale la pena tener paciencia.

En particular, valió la pena para Englert y Higgs, quienes recibieron
el Premio Nobel en 2013 por su predicción correcta \cite{Nobel13};
tristemente, Brout había muerto poco antes, en 2011.
En abril de 2024 nos llegó la noticia del fallecimiento de Higgs,
a los 94 años, después de una breve enfermedad.

\begin{figure}
\centering
\includegraphics[scale=.3]{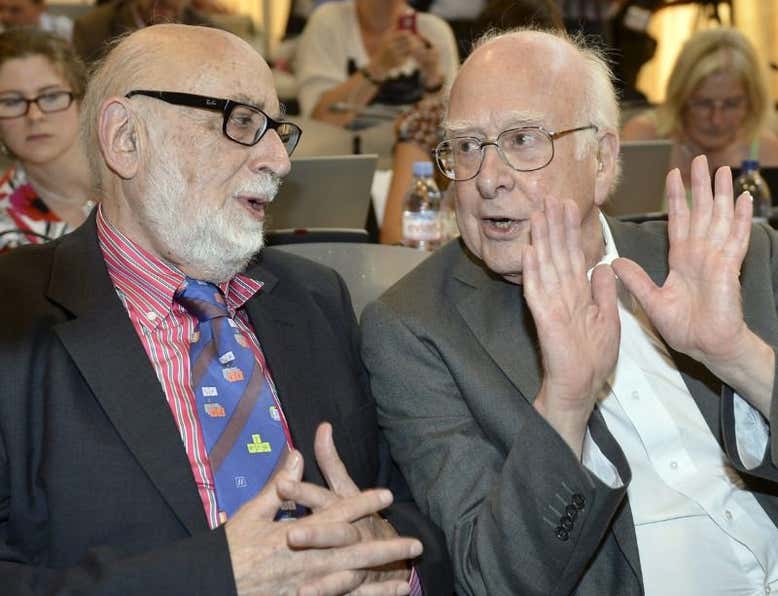}
\caption{El 4 de julio de 2012 el CERN hizo público el
descubrimiento del bosón de Higgs. Peter Higgs conmovido hasta
las lágrimas en la ceremonia dijo: ``Felicitaciones a todos los
involucrados en este logro tremendo. Para mí es algo verdaderamente
increíble que esto pasó durante mi tiempo de vida.''
La Real Academia Sueca de las Ciencias otorgó el premio Nobel
de Física en 2013 a Fran\c{c}ois Englert (izquierda) y a Peter
Higgs (derecha) por ``el descubrimiento teórico de un mecanismo
que contribuye a nuestra comprensión del origen de la masa de
las partículas subatómicas $\dots$'' \cite{YouTube}.}
\end{figure}

\section{Mecanismo de Higgs}

Hasta donde sabemos, el mundo consiste de partículas elementales, que
son indivisibles, y existen en pocos tipos (decimos 25, pero depende
un poco de cómo se cuenta). Ejemplos famosos son el electrón y el
fotón (la partícula de la luz).

Su descripción relativista funciona con ``campos'', cantidades
abstractas, presentes en todo el universo, en cualquier momento.
En un punto, un campo puede tomar diferentes estados, que dependen
del tiempo. Si los campos en una región están en su estado base,
percibimos el vacío. Las excitaciones son cuantizadas y se
manifiestan como partículas elementales -- esta es la idea de la
{\it teoría cuántica de campos.}

Existe un campo para cada tipo de partícula elemental,
y sus excitaciones pueden moverse (como ondas), interactuar,
generar y destruir partículas (esto es un requisito para
la compatibilidad con la Relatividad Especial, que falta en
la Mecánica Cuántica).\footnote{Ref. \cite{WBparti} presenta
  otra descripción divulgativa, pero más extensa, de
  partículas elementales.}

Un concepto central son las simetrías: una simetría significa
la invarianza de las propiedades físicas bajo un grupo de
transformaciones de uno o varios campos. Distinguimos simetrías
{\em globales} y {\em locales:}

\begin{itemize}

\item En una simetría {\em global,} un campo se transforma de la
  misma manera en todas partes. Se puede imaginar un grupo de personas
  que hacen gimnasia colectiva, todas hacen el mismo movimiento,
  puede ser sincronizado con música.
  (La imagen es un poco simplista porque los campos se transforman
  de la misma manera incluso en todo el espacio-tiempo.)
  
\item El caso de una simetría {\em local} se puede imaginar como
  gimnasia caótica: cada persona se mueve como quiere, de manera
  independiente. Esto significa que los campos pueden ser transformados
  independientemente en cada punto del espacio-tiempo.

  Es claro que este tipo de simetría permite muchas más
  transformaciones.
  Lograr una simetría local es más difícil, pero conduce
  a restricciones más fuertes, y por lo tanto a una poderosa
  capacidad de hacer predicciones.
  
  Técnicamente, se introduce un campo adicional, conocido como
  {\em campo de norma,} que transforma de manera que compensa el cambio
  relativo entre puntos cercanos en una transformación simultánea.
  Este concepto exitoso describe la transmisión de interacciones,
  pero solamente funciona si la simetría local es exacta.
  
\end{itemize}

Una categoría importante de partículas es conocida como
``fermiones'': los fermiones elementales (conocidos) tienen
``espín $1/2 $'' en unidades naturales.\footnote{Para
usar unidades naturales, se coloca la constante
cuántica de Planck y la velocidad de la luz en el vacío a 1,
$\hbar = c =1$.}
El espín es un grado de libertad interno que se manifesta como momento
angular. Los fermiones del Modelo Estándar son el electrón, sus dos
``primos'' más pesados (el muón y el tauón), los neutrinos (mucho
más ligeros y sin carga eléctrica)\footnote{El conjunto del
electrón, muón, tauón y los tres neutrinos es conocido como
los {\em leptones}.}
y los cuarks (constituyentes
de partículas compuestas, como el protón y el neutrón).

Un fermión puede existir en dos variantes, con ``quiralidad''
izquierda o derecha; se puede imaginar como manos,\footnote{Efectivamente,
el término viene de ``kheir'', que significa ``mano'' en griego.} 
o guantes, izquierdo o derecho, pero en un sentido abstracto.

\begin{figure}
\centering
\includegraphics[scale=1]{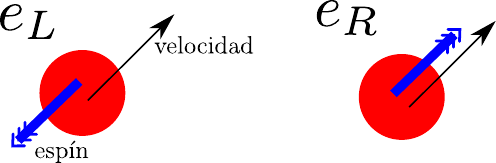}
\caption{Representación simbólica de un electrón sin masa con
quiralidad izquierda $e_L$ y quiralidad derecha $e_R$. La dirección
del movimiento es indicada por la flecha delgada y la dirección del
espín por la flecha gruesa. Para quiralidad izquierda, la velocidad y
el espín apuntan en direcciones contrarias, pero para la quiralidad
derecha apuntan en la misma dirección.}
\label{electron}
\end{figure}

Supongamos por ejemplo un electrón sin masa: en este caso,
el electrón izquierdo ($e_L$) y derecho ($e_R$) son independientes,
y su espín apunta en contra a (para $e_L$) o en (para $e_R$) la
dirección de su movimiento (una particula sin masa no
puede estar en reposo). Esto es ilustrado simbólicamente en
la Figura \ref{electron}.

Incluir un término de masa en la teoría requiere de un producto de los
campos del electrón izquierdo y derecho (se puede imaginar que las dos
manos se agarran). Entonces ya no son independientes, y bajo una
simetría tienen que transformarse de la misma manera.

Sin embargo, este no es el caso en la teoría electrodébil de Glashow
\cite{Glashow}: esta teoría permite, por ejemplo, transformaciones
locales (``de norma'') que solamente afectan al $e_L$, pero no al $e_R$.
Aquí estaba el problema: dicha teoría parecía ser incompatible con el
término de masa del electrón (y de otros fermiones), pero sabemos
que el electrón sí tiene una masa de $M_e \simeq 0.511\ {\rm MeV}$
(todavía en unidades naturales).

De hecho, la situación era aún peor. Las partículas de norma
que transmiten la fuerza débil se llaman $W^{\pm}$, $Z^0$
(con carga eléctrica $\pm1, \, 0$), por ejemplo los $W$ son
responsables del decaimiento radioactivo. Esta fuerza tiene un
alcance muy corto (como $10^{-17}$~m) que solamente se puede
explicar si $W^{\pm}$, $Z^0$ tienen masas grandes (están entre
las partículas elementales más pesadas que conocemos, con
masas de $M_W = 80.4 \, {\rm GeV}$ y $M_Z = 91.2\,{\rm GeV}$). 
Pero igual que la masa del electrón, parece que la simetría
de norma -- que tiene que ser exacta -- requiere $m_W = m_Z = 0$.

El acertijo de dónde pueden venir estas masas de partículas
de norma era una ``pregunta matadora'' con la cual Wolfgang Pauli
arruinó un seminario de Chen-Ning Yang en Princeton en 1953
sobre teorías de
norma con un grupo de simetría no abeliano (ahora conocidas como
teorías de Yang-Mills). Sin conocer el mecanismo de Higgs, Yang no
logró contestar, pero Pauli insistió tanto hasta que Yang se
sentó, frustrado. Finalmente Robert Oppenheimer tuvo
que animarlo para continuar su charla \cite{Shifman}.

Entonces, ?`cómo funciona la salvación de esta teoría,
el mecanismo de Higgs? Primero, se agrega otro campo más,
el {\em campo de Higgs}, usamos la notación $\phi (x)$.
La variable $x$ es un punto del espacio-tiempo, y $\phi$ es
un campo escalar, sus fluctuaciones representan partículas 
con espín 0. Para establecer un término de masa del electrón,
ahora se forma un producto de {\em tres} campos, $e_L$, $\phi$ y
$e_R$.\footnote{Realmente necesitamos parcialmente anti-campos,
que representan anti-partículas, pero ignoramos
este aspecto en el contexto de este artículo de divulgación.}
El campo de Higgs también transforma bajo la simetría local,
de tal manera que el término en su totalidad sí es invariante
de norma.

Entonces de esta manera se puede agregar un término permitido
(invariante de norma), pero ?`esto proporciona una masa al electrón?
Posiblemente sí, puede funcionar con el escenario siguiente.

\begin{figure}
\vspace*{-5mm}
	\begin{center}
	\includegraphics[scale=2]{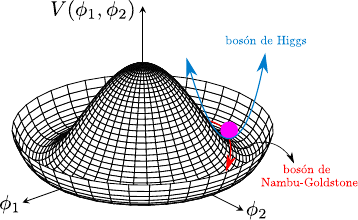}
	\end{center}
	\caption{Potencial de Higgs: desde la perspectiva de la
        cima el potencial es simétrico bajo rotaciones.
        Sin embargo, desde la perspectiva de la bola el potencial
        parece no presentar dicha simetría, esto es el ``rompimiento
        espontáneo de simetría''. Las fluctuaciones {\em tangenciales}
        del campo a lo largo del círculo de mínimos se manifiestan
        como un bosón de Nambu-Goldstone si la simetría es global.
        Si la simetría es local, este bosón de Nambu-Goldstone
        es ``comido'' por un campo de norma que adquiere masa. Las
        fluctuaciones {\em radiales}, perpendiculares al círculo de
        mínimos, se manifiestan como una partícula masiva. Si
        tratamos con el potencial de Higgs como ocurre en el Modelo
        Estándar, esta partícula masiva es el famoso bosón de
        Higgs.}
\label{sombrero}
\vspace*{-2mm}
\end{figure}

A bajas energías, el campo de Higgs ``se congela'' en su estado
base, se manifiesta casi como una constante. Esta constante no
tiene que ser cero: realmente $\phi$ tiene 4 componentes reales, pero
nos podemos imaginar que sean 2 solamente,
$\phi_1 , \, \phi_2 \in \R$, que parametrizan un plano.
Este campo viene con un potencial $V(\phi_1, \phi_2)$ de la forma
de un sombrero, con el valor cero al centro, pero hay un anillo
de mínimos que corresponden a un valor
$|\phi|^2 = \phi_1^2 + \phi_2^2 > 0$, como está ilustrado en la
Figura \ref{sombrero}.
Este ``valor esperado en el vacío'' toma el papel de la masa
del electrón que el modelo necesita (hasta un coeficiente),
$M_e \propto |\phi|$, y de manera análoga también se obtienen
las masas $M_W$ y $M_Z$, todo conforme a la simetría de norma
(regresaremos a este tema).

Parece todo bien, pero hay otro problema todavía, y a esto
Coleman se refirió en su comentario sobre el seminario de
Higgs en Harvard, que hemos mencionado en la Sección 1.

El potencial sombrero tiene una simetría bajo rotaciones por
el eje central. Suponemos que el campo $\phi$ elige uno de
los mínimos: el proceso de esta elección se denota como
``rompimiento espontáneo de la simetría'': desde
la perspectiva de un mínimo específico, ya no se ve la
simetría de rotación. En Ref.\ \cite{HiggsBol} hemos
descrito este proceso con la analogía del {\em Asno de Buridan,}
que está sediento y rodeado por un abrevadero de agua, pero
tiene que decidirse en que dirección camina para beber agua.

Pequeñas fluctuaciones del campo más allá de su estado
mínimo corresponden a partículas. Si una fluctuación es
{\em radial,} cuesta energía porque el potencial sube -- esto
es una partícula masiva (la curvatura del potencial en la
dirección radial corresponde a su masa al cuadrado).
Por otro lado, una fluctuación {\em tangencial} no necesita
energía, pues el campo se queda con energía mínima.
Esto es un ejemplo de una partícula sin masa, conocida
como un {\em bosón de Nambu-Goldstone} \cite{Nambu,Goldstone}.
Según el Teorema de Goldstone \cite{GSW}, estos bosones aparecen
cuando una simetría continua (como la rotación en este
ejemplo) se rompe espontáneamente.

Si el mecanismo funciona como se ha descrito antes, se queda
la pregunta: ?`dónde está este bosón de Nambu-Goldstone?
Por ser sin masa, tendría que tomar un papel importante
y dominar la física a bajas energías (donde partículas
muy pesadas no se manifiestan). Pero ninguna partícula de
este estilo fue observada. Entonces para justificar el mecanismo,
se tiene que ``evadir el Teorema de Goldstone'', como dijo
Coleman, pero los físicos en Harvard dudaron si esto era posible.
No eran los únicos, por ejemplo Klaus Hepp, prominente
físico matemático, advirtió a Higgs que esto no podría
funcionar porque el Teorema estaba demostrado con álgebra
C$^{*}$, un formalismo que Higgs no conocía, pero él expresó
sus dudas de las suposiciones en esta demostración \cite{boson}.

Ahora sabemos que el teorema sí es correcto, pero solamente
se refiere al rompimiento de una simetría continua
{\em global} -- esta suposición estaba escondida.
La observación crucial era que la situación es diferente
en el caso de una simetría {\em local}: en este caso, los
mínimos están conectados por transformaciones locales, o
transformaciones de norma, por lo tanto físicamente idénticas.
Así no hay fluctuaciones físicas entre los mínimos, y no hay
bosones de Nambu-Goldstone. Lo que pasa, y completa el mecanismo
de Higgs, es que el bosón de norma adquiere masa, lo que significa
que el grado de libertad que tenía el bosón de Nambu-Goldstone
se convierte en un grado de libertad longitudinal del bosón de
norma (sin masa solamente tiene grados de libertad transversales).
En lenguaje popular, se dice que el bosón de norma ``se come''
al bosón de Nambu-Goldstone: este último ya no está, pero
el primero se hace ``gordo''.

Esto había observado Anderson antes en un típico superconductor:
en su interior, a muy baja temperatura, el fotón adquiere masa,
por esto casi no puede penetrar al superconductor -- un fenómeno
conocido como {\em efecto de Meissner-Ochsenfeld.}
Brout, Englert y Higgs extendieron este efecto a modelos
relativistas \cite{EB,Higgs}, y Weinberg y Salam a la fenomenología
de la interacción electrodébil \cite{Weinberg,Salam}.
Hemos mencionado que el campo de
Higgs tiene 4 componentes reales: siempre hay una fluctuación
masiva radial, y entonces 3 bosones de Nambu-Goldstone si tratamos
con una simetría global. Cuando la promovemos a una simetría
local, los bosones $W^+$, $W^-$ y $Z^0$ ``se comen'' a estos
bosones de Nambu-Goldstone y adquieren masas,
mientras que el fotón se queda sin masa (en circunstancias
normales) y describe el electromagnetismo que tiene alcance largo.

Ya hemos visto que el mecanismo también proporciona una masa al
electrón, y de la misma manera aplica al muón, tauón y a
todos los cuarks. ?`Y qué pasa con el campo de Higgs?
Sabemos que 3 de sus 4 componentes serían bosones de
Nambu-Goldstone que desaparecen, pero está la cuarta componente
todavía, que corresponde a la fluctuación radial, es decir,
a una partícula masiva. Su existencia es una predicción del
mecanismo de Higgs, y en este aspecto el segundo artículo de Higgs
de 1964 era algo más explícito que los otros trabajos originales.
Esto era en el contexto de modelos juguete todavía, pero una vez se
aplicó a un modelo fenomenológico, se concluyó que dicha
{\em partícula de Higgs} tendría que ser {\em observable.}

La teoría no predice la masa de la partícula de Higgs, $M_{\rm H}$
(se pueden derivar cotas solamente que eran tema de discusión durante
muchos años), por lo que la búsqueda experimental fue difícil.
Al inicio del siglo XXI, los experimentos del
Gran Colisionador de Leptones y Protones (LEP) del CERN demostraron
que tiene que tener una masa $M_{\rm H} > 114\, {\rm GeV}$. Entonces se
sabía que tiene que ser muy pesado (si existe), tal que su creación
requiere colisiones de altas energías. Esto también implica que
su tiempo de vida es muy corto, decae en promedio en $10^{-22}$
segundos, y no puede dejar trazas en ningún detector.

Un experimento tiene que capturar los productos de su decaimiento
que permiten la reconstrucción de la partícula de Higgs como
estado intermediario, o ``resonancia'' -- por muy poco tiempo --
en una colisión a
altas energías. El análisis de las partículas que resultan al
final del proceso permiten reconstruir las propiedades de la
partícula de Higgs, en particular su masa de $M_{\rm H} = 125\,
{\rm GeV}$ y su espín 0, que confirma que se trata de una partícula
escalar.

En el canal más limpio que fue observado, el decaimiento de la
partícula de Higgs termina con dos fotones, un estado final que
no sería posible si, por ejemplo, la partícula original
tuviese espín 1, como Landau había demostrado \cite{Landau}.
Pero los experimentos ATLAS y CMS estudiaron (independientemente)
muchos más decaimientos en gran detalle -- como, por ejemplo,
con un estado final de cuatro leptones -- y no dejan ninguna duda
de la existencia de la partícula de Higgs, y del mecanismo
correspondiente.

\begin{figure}
\centering
\includegraphics[scale=0.2]{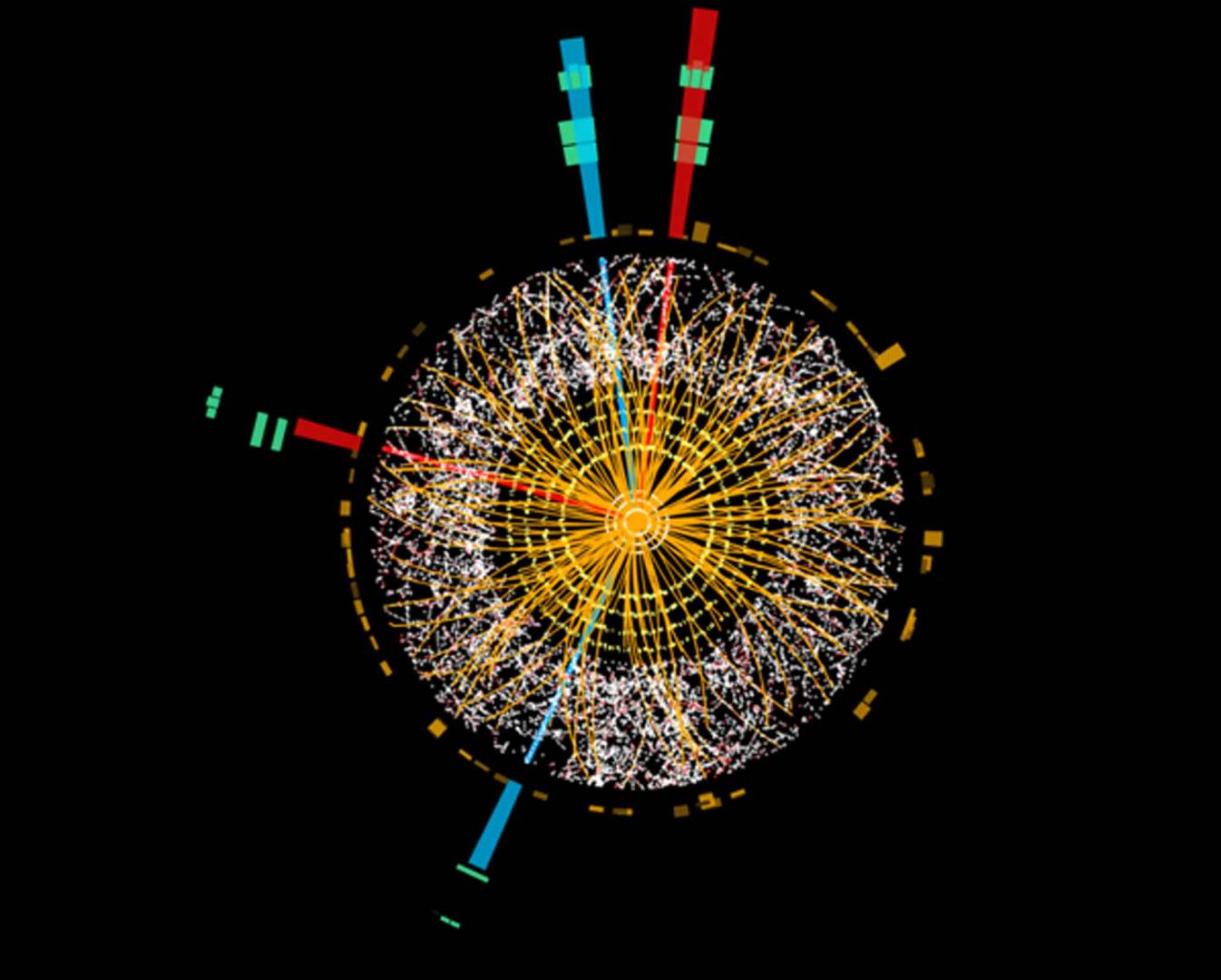}
\caption{Canales posibles del decaimiento del bosón de
Higgs cuando decae a dos bosones $Z$ que a su vez decaen cada
uno en un par lepton-antilepton. 
En la imagen observamos 4 leptones (líneas rojas y azules) que
posiblemente hayan sido producidas por el decaimiento del bosón de Higgs.}
\end{figure}

\begin{figure}
\centering
\includegraphics[scale=0.4]{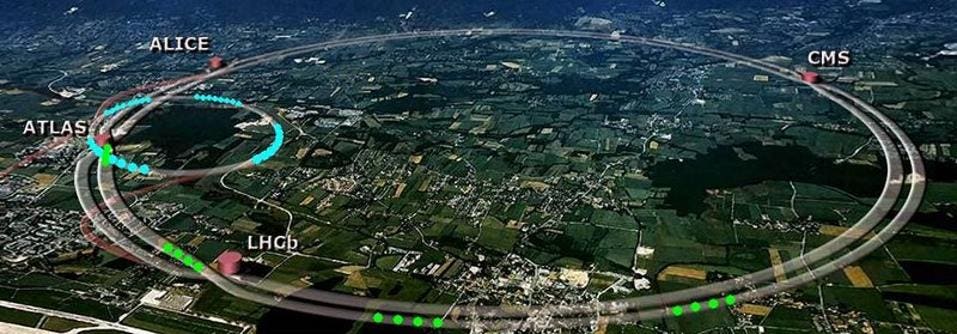}
\caption{El Gran Colisionador de Hadrones (LHC por sus siglas en
inglés) se encuentra ubicado en la región fronteriza entre
Francia y Suiza.
Sus cuatro experimentos principales se llaman ATLAS y CMS (que
de manera independiente encontraron evidencia de la existencia
del bosón de Higgs), ALICE -- con participación mexicana --
y LHCb. Los otros países latinoamericanos involucrados en
estas colaboraciones son \cite{CERN}: Argentina, Brasil, Chile,
Colombia, Costa Rica, Cuba, Ecuador y Perú.}
\vspace*{-2mm}
\end{figure}

Esto representa un éxito espectacular de la física de partículas
elementales. Aún así, no todo está resuelto todavía. Desde la
perspectiva conceptual, se queda el {\em problema de la jerarquía}
y respecto a la fenomenología, el mecanismo de Higgs no explica
el origen de todas las masas que obervamos. Terminamos con
comentarios breves sobre estos asuntos:

\begin{itemize}

\item {\em Problema de la jerarquía:} Si empezamos con un sistema
clásico (sin efectos cuánticos) y suponemos una masa $m_{\rm H}^{(0)}$
en la magnitud de las masas de otras partículas,
es natural que las correcciones cuánticas suban dicha masa
drásticamente, a un valor $m_{\rm H}$ del orden de la ``escala de
Planck'' (determinada por la constante de la gravitación).
Pero esto conduce a un valor de $m_{\rm H}$ muy alto, típicamente
como $10^{17}$ veces su valor observado.

Lo que se podría hacer es suponer un valor de $m_{\rm H}^{(0)}$
extremadamente negativo, así que el efecto cuántico
se cancela casi totalmente, y se queda un resto diminuto
de 125\,GeV. Pero esta aproximación -- con una cancelación
entre dos contribuciones tremendas que deja un resto diminuto --
no parece natural. Esto es conocido como el ``problema de la
jerarquía''. Sin embargo, no es una paradoja, se puede llegar
de manera consistente a 125\,GeV, y la pregunta de qué tan
grave es este problema es un poco filosófica.
El salto jerárquico de energía también depende del esquema de
regularización que se aplique. 

\item Los {\em neutrinos} (descritos con el símbolo $\nu$) tienen
un papel especial: en la forma tradicional del Modelo
Estándar se supuso que tenían masa cero, $M_{\nu} =0$,
y que solamente existía el neutrino con quiralidad izquierda,
$\nu_L$.

Esto es consistente en teoría, pero a finales del siglo XX
se observó que los neutrinos sí tienen una pequeña masa,
$m_{\nu}>0$ -- repetimos que Ref.\ \cite{neutrinos} presenta una
revisión semi-divulgativa del tema.

De primera vista, conforme con nuestra descripción anterior,
parece inevitable que exista el neutrino derecho, $\nu_R$.  
Esto permite la aplicación del mecanismo de Higgs para los
neutrinos, y además otro tipo de masa, solamente para el
$\nu_R$, conocido como ``masa de Majorana''.

Sin embargo, el $\nu_R$ no es observado, y su existencia no
es realmente inevitable: podemos construir un término de masa
que involucra únicamente el campo $\nu_L$ \cite{Weinberg79}.
Este término no es
renormalizable, pero hemos mencionado en la Sección 1 que ya no
se da tanta importancia a esta propiedad. Entonces este escenario
sería la alternativa, en el marco del Modelo Estándar interpretado
como teoría efectiva que funciona en cierto rango energético.

\item Hasta este punto, el mecanismo de Higgs explica las masas
de las partículas elementales (con la posible excepción de
los neutrinos), y hay otras partículas como el fotón que se
quedan sin masa. Parece una imagen completa del origen de la masa.

Sin embargo, el mundo real es diferente: en realidad, la masa de un
objeto macroscópico de nuestra vida cotidiana viene solamente
por $1 \dots 2 \, \%$ del mecanismo de Higgs, que conduce a las
masas de las cuarks.

Estas masas cotidianas consisten principalmente de masas de
nucleones (protones y neutrones) que por su parte consisten
esencialmente de energía de {\em gluones} (otras partículas de
norma, que transmiten la interacción fuerte): tienen masa cero,
pero son confinados al interior de un nucleón (u otra partícula
compuesta por la interacción fuerte). Su energía se
manifiesta como casi toda la masa del nucleón, mientras que
las masas de los cuarks solamente proporcionan dicha contribución
al nivel de $\sim 1 \dots 2 \, \%$.

El interior de un nucleón es un sistema muy, pero muy complejo;
por mucho tiempo pareció imposible calcular algo conclusivo
al respecto. Sin embargo, hace poco más de una década 
se logró calcular por ejemplo la masa del nucleón,
$M_N \simeq 939\,{\rm MeV}$, de primeros principios, hasta una
incertidumbre del orden de 1\,\%. Este cálculo captura
el revoltijo hiper complicado de gluones (y ``cuarks de mar'',
parejas inestables de un cuark y su anti-cuark), y el resultado
es compatible con los experimentos. La pregunta de cómo estos
cálculos son posibles sería tema para otro artículo $\dots$

\end{itemize}

\noindent
{\bf Agradecimientos:} Agradecemos a Yeslaine Hernández-Hernández,
Víctor Muñoz-Vitelly y Jaime Fabián Nieto Castellanos por revisar
el manuscrito.
Este trabajo fue apoyado por {\it UNAM-DGAPA} a través del proyecto
PAPIIT IG100322, y por el {\em Consejo Nacional
de Humanidades, Ciencia y Tecnología} (CONAHCYT).

\end{document}